# The Joule Point: an Energy-Optimal Operating Point for AI Inference

**Alexander Apartsin[1] · Yehudit Aperstein[2]**

[1]*Holon Institute of Technology, Holon, Israel* [2]*Afeka Academic College of Engineering, Tel Aviv, Israel*

**Abstract**

Data centers serving AI inference strive to maximize GPU utilization, running their cards at full power by default: this maximizes throughput and holds latencies down, but it is energy-inefficient, spending substantially more energy per inference than the same work needs at a better operating point. That inefficiency is a scheduling choice, not a hardware limit, and it becomes visible only when we stop counting GPUs and start counting joules. The cause is physical: for a given workload, a GPU's board power rises superlinearly along its power-performance curve, on top of a fixed power floor that is paid for as long as the job runs, so energy per inference is U-shaped in the operating point (GPU, power cap), a function of that point rather than of the rate alone. The minimum, which we name the *Joule Point*, is a cap at 43 to 46 per cent of a large GPU's rated power; capping to it cuts energy per inference by roughly a quarter to a third at a modest, exactly priced cost in capital and latency: each request runs about 1.2 times slower, and holding aggregate throughput takes that same factor more cards (the card multiplier equals the latency ratio). Under load the Joule Point is nearly a per-card-type constant, so a single static cap per card type, set once, captures nearly all the saving (mean penalty under one per cent), turning the online per-job search that prior systems run into a one-time characterization. We ground this in ELF, a dense power-cap dataset that sweeps 20 inference models across four GPUs, three of them finely enough within each workload-GPU pair to locate the Joule Point and fit the law; replaying it in a fleet simulation under a power budget, capping each job to the least power meeting its deadline spends 18 to 45 per cent less energy per served job, a gain that comes from each job's own operating point, not the hardware mix. These results recast data-center energy as a schedulable resource an operator can adapt to a budget, an electricity price, or a carbon signal while meeting its service targets.

## 1. Introduction

Inference dominates the lifetime compute of a widely deployed AI model, and the electricity it draws has become a leading constraint on how fast data centers can grow. The usual practice is to run each GPU at its full rated power (its thermal design power, TDP) and to add more GPUs when more capacity is needed. This paper argues that this default wastes energy, that the waste is invisible in the terms schedulers reason about today, and that changing those terms reveals both the waste and a family of better decisions.

The trade-off is familiar from transport: ships slow-steam and aircraft cruise at an economical speed rather than a maximum one, because fuel burn rises superlinearly with speed and the cheapest trip balances that burn against time and schedule [59, 60]. The Joule Point is the computing analogue, trading power for time to minimize energy per inference within a latency budget.

What actually costs money and carbon is the energy spent per unit of useful work, the joules per inference. Yet scheduling in hardware space, choosing which GPU runs a job and how many GPUs to use, never treats the power a GPU draws as something to decide: running uncapped is an unspoken default rather than a deliberate choice. We show that on large GPUs, holding a GPU to its energy-optimal power cap cuts the energy per inference by 29 to 31 per cent (Section 6) at a small, predictable cost in latency, and that in rented-GPU settings the party who benefits from running at full power is not the party who pays the electricity bill (Section 8).

Our remedy is to describe every run, for a given workload, by its *operating point*, the pair (GPU g, power cap p), and to read every decision off a measured response surface and its slopes: how throughput responds to power, and where the energy per unit of work is lowest. We call this *energy space*. In these terms the good decisions become simple calculations and the bad ones become visible mistakes. We ground the argument in measurement rather than analytical modeling, releasing ELF (Section 4), a dense power-cap sweep for 20 inference models across four GPUs.

**Contributions.** Concretely, the paper releases a measurement dataset, establishes the response law and its Joule Point, turns them into fleet-scheduling decisions, and locates the economic obstacle to adoption:

1. **ELF, a released measurement dataset** (Section 4): dense power-cap sweeps for 20 inference models on four GPUs, with board power, throughput, and latency at each operating point, the surface that makes energy-space analysis possible without new hardware runs.
2. **A response law and a coordinate justification** (Section 5): board power follows $P(\theta) = P_0 + a\theta^\beta$ (normalized rate $\theta$, fitted floor $P_0$, scale coefficient $a$, exponent $\beta$) at median $R^2 = 0.99$, and it is a law of the operating point, not the rate alone: pool measurements across batch sizes and the single-curve fit falls apart. This is the empirical license for scheduling in energy space rather than hardware space.
3. **The Joule Point and its economics of use** (Section 6): an energy-optimal cap at 43 to 46 per cent of TDP on large GPUs, an exact cost identity (the card multiplier to hold throughput equals the latency ratio), and the finding that placing the cap is a one-time characterization, not an online search: under load the Joule Point is almost model-invariant, a per-card constant, so one fixed cap per card serves all twenty measured workloads (a mean penalty of 0.9 per cent on the A100 and 0.4 per cent on the A10G versus each model's own optimum), where prior energy-aware systems search for this operating point per job at runtime. The exponent $\beta$ lives in the measured run, not in a model's spec sheet, which is what makes these one-measurement procedures the right way to place the cap.
4. **A fleet simulation** (Section 7): a trace-driven scheduler that serves equal work for 18 to 45 per cent less energy under a power budget; the saving comes from capping each job down to its service-level objective (SLO), a property of the operating point: even a fleet of identical GPUs captures it in full.
5. **Operator implications and an incentive analysis** (Section 8): how to size, cap, and price a fleet in energy space, and the finding that per-hour billing makes the renter's cost optimum the energy-worst operating point, identifying the central obstacle to adoption as economic rather than technical.

Section 2 positions this against prior work on energy-proportional computing, GPU power management, and adaptive inference. Section 3 defines the energy-space coordinate system. Sections 4 to 6 develop the dataset, the law, and the Joule Point; Section 7 tests them in a fleet simulation. Section 8 draws out implications for operators, and Section 9 gives limitations and Section 10 concludes.

## 2. Related work

**Energy-proportional computing.** Barroso and Holzle [1] argued that servers should draw power in proportion to the work they do, and that the fixed, work-independent share of power is the central inefficiency. The fixed floor $P_0$ in our law is exactly that share measured for modern GPUs, and the Joule Point is where paying it stops being worthwhile. At facility scale the same fixed share drives power over-provisioning [15], the datacenter analogue of running a single GPU below its floor.

**Convex power-performance and speed scaling.** That energy is minimized at a sub-maximal speed is a classical result. Speed-scaling theory models power as a convex function of processing rate and schedules to a deadline: Yao, Demers, and Shenker [35] analyze minimum-energy scheduling under a power law $P(s) = s^p$, Miyoshi et al. [36] give a criterion for an energy-optimal operating point (the closest ancestor of our Joule Point), and Gandhi et al. [37] show that maximum power is not optimal under a farm-wide budget. The theory supplies the existence of a sub-maximal optimum; ELF supplies its measured form for GPU inference: ELF establishes $P(\theta) = P_0 + a\theta^{\beta}$ along fixed operating-point families with a nonzero floor, and shows that pooling different (batch, cap) points destroys any rate-only model (the pooled fit collapses to a median $R^2$ near zero), which is what forces the coordinate change from rate to operating point.

**GPU power capping and DVFS.** That a GPU has an energy-optimal clock or power limit below its maximum is well established. Tang et al. [17] and the survey of Mei et al. [18] measure energy savings from GPU DVFS across deep-learning workloads, Zeus [2] builds an online optimizer over the batch-size and power-limit knobs for training, BatchDVFS [21] couples batch size with frequency under a cap, and Perseus [20] and μ-Serve [22] remove energy bloat, power drawn without a matching gain in throughput, in training and in serving through frequency scaling; the energy-delay product these methods trade against goes back to Gonzalez and Horowitz [19], and the limits of the underlying knob are mapped by architectural power models (Hong and Kim [38], GPUWattch [39]) and by direct GPU undervolting studies [40]. We take that sub-TDP optimum as settled ground and build on it in three ways: a released dense response *surface* across 20 models and four GPUs rather than a per-deployment optimizer; the operating-point law with its closed-form Joule Point (Section 6) and its "power is not a function of rate" corollary, which motivates the coordinate change itself; and an explicit accounting of the economic incentive that has kept the optimum unused in practice.

**Power- and energy-aware LLM serving.** A large systems literature raises throughput and meets latency targets by scheduling in hardware space, deciding which GPUs run what: iteration-level batching in Orca [23], paged attention in vLLM [24], chunked prefill in Sarathi-Serve [25], prefill and decode disaggregation in DistServe [26] and Splitwise [16], and, at cluster scale, heterogeneity-aware assignment of jobs to GPU types in Gavel [27]. A closer group manages power directly: POLCA [3] oversubscribes inference power, DynamoLLM [4] and throttLL'eM [45] scale frequency to meet SLOs, GreenLLM [46] scales prefill and decode frequencies separately, TAPAS [47] schedules under thermal and power limits, a serverless-serving scheduler [44] co-controls placement and GPU operating points, and PALS [43] tunes the power cap jointly with batch size inside vLLM, actuating the same knob we study. These controllers treat the energy-optimal operating point as an online, per-job control target, searched for at runtime. We measure that target directly and find that under sustained load it is nearly a per-card constant (Section 6), which turns an online control problem into a lookup; online control keeps its value where load or the prefill/decode phase mix varies enough to move that floor. The layer beneath these systems is thus a coordinate system and a calculus over power-cap operating points, the response law and its Joule Point across many models and GPUs (Sections 5 and 6), on which any such controller can act.

**Adaptive and selective inference.** A separate line adapts the amount of computation to each input: early-exit networks [5] and our own CalexNet [11], multi-scale budgeted architectures [54], conditional and dynamic routing per input [6, 7, 55], early exiting for transformers [53], and speculative or self-speculative decoding [8, 52]. These methods save by doing less computation per input; the power cap saves by spending less energy per unit of computation. The two axes are conceptually distinct and can compose, and Section 10 proposes unifying them in a single energy-space design, with the thermodynamics of prediction [9] and Landauer's limit [10] setting the ultimate floor for that energy-per-information view.

**Measuring inference energy, and released datasets.** Public efforts increasingly quantify inference energy: MLPerf Power [31] standardizes system-level energy benchmarking, the ML.ENERGY benchmark [32] automates inference-energy measurement across dozens of models, From Words to Watts [33] profiles LLM inference across GPU configurations, Watt Counts [48] compares 50 LLMs across ten GPUs for energy-optimal hardware selection, BUTTER-E [49] releases tens of thousands of DNN energy runs, MELODI [50] profiles energy across prompts and frameworks, and de Vries [34] projects the sector's electricity growth. These are broad across models, prompts, or hardware, and report energy at whatever operating points a deployment happens to use. ELF is instead dense *within* each workload-GPU pair, sweeping the power cap crossed with batch size over 20 models and four GPU classes, which is what makes the local slopes, the compute-power elasticity, and the Joule Point measurable.

**Carbon- and grid-aware data centers.** A parallel line makes the whole facility responsive to the grid: a 256-GPU cluster has been field-tested shedding about a quarter of its power for hours [12], schedulers shift load in time toward cleaner electricity [13] or across regions toward cheaper power [41], renewable-aware batch scheduling defers jobs to sunny hours [42]. Holistic frameworks co-design renewables, batteries, and scheduling [28], virtualized energy systems expose solar and battery state to applications [29], elastic workloads scale with carbon intensity [30], carbon-aware inference trades placement and quality for emissions [51], and the view of data centers as deferrable demand-response load is long-standing [14]. All of this treats the facility as the unit of control. The operating points we measure turn the same shedding into a per-job, per-SLO decision, priced by the carbon and demand-response signals that give a watt its value (Section 8).

**Energy-aware pricing and split incentives.** Charging for compute by its energy rather than its wall-clock goes back at least a decade: tenant-facing flexibility pricing has been used to extend demand response into cloud data centers [56], and energy-aware cost models price virtual machines by attributed energy [57, 58]. Zhan et al. [56] in particular name the split incentive between an operator that wants flexibility and tenants billed by conventional usage. Our contribution is to locate that split incentive at the GPU operating point and quantify it: under per-hour rental the renter's throughput-maximizing point and the operator's energy-minimizing Joule

Point diverge, and at today's rental-to-electricity price ratios a simple per-joule pass-through is too weak to close the gap (Section 8).

## 3. Energy space and its calculus

An execution is fixed by its *workload state* $w$ (the model, batch size, precision, and serving stack) together with an *operating point* on a GPU: the pair (GPU g, power cap p). Conditional on $w$, the cap induces an achieved inference rate (batch forward passes per second) and board power P; where the cap binds, $P = p$, and we verify that measured draw tracks the cap across the swept range. We write this rate in normalized form as $\theta$, the achieved rate as a fraction of that workload's own maximum rate on the same card ($\theta \in (0,1]$, with $\theta = 1$ at full power); writing the achieved throughput as $R$ and its ceiling as $R_{\max}$, $\theta = R/R_{\max}$. Normalizing this way makes the response law comparable across workloads and GPUs, which is what lets a single law hold per card (Section 5). The rate $\theta$ is a rate of *work* (inferences per unit time), not of energy; the rate of energy is the power P. The response curves $P(\theta)$ below are therefore per-workload-state families, not a single function of rate; that non-uniqueness is what forces the coordinate change from rate to the operating point. Two quantities read off this surface carry the decisions in the rest of the paper. Energy per inference $E = P/\theta$ has an interior minimum where its derivative vanishes, $dE/d\theta = 0$: the fixed floor is amortized over a faster job while the superlinear tail is paid per inference, and the two balance at a sub-maximal power. The service-level-objective (SLO) power $P_{\text{SLO}}$ is the least power that still meets a latency target. Every decision below is a statement about these quantities: Sections 5 and 6 measure the response curves and locate the minimum, Section 7 moves each job toward $P_{\text{SLO}}$ under a fleet budget, and Section 8 weighs the resulting cost, capacity, and latency trade-offs for an operator. A representation limited to hardware placement exposes none of them.

## 4. The ELF dataset

Every decision in this paper reads off a measured surface, so we release that surface as **ELF**, the Energy-Latency Frontier dataset. ELF is a dense power-cap sweep, from each GPU's driver floor to its TDP (eight cap levels per card), for 20 inference models on four NVIDIA GPUs.

**Workloads.** The 20 models span the inference mix: an autoregressive LLM decode step with a pre-filled KV cache (llm_decode), a Stable-Diffusion UNet forward (sd_unet), three vision transformers (ViT-B/16, ViT-B/32, Swin-T), and fifteen convolutional networks, all at FP16. **Hardware.** The analysis covers three classes spanning a fivefold power range: L4 (72 W TDP), A10G (300 W), and A100-SXM4 (400 W). The release also includes a fourth card, the T4 (70 W), which we exclude from the quantitative analysis as a boundary case (its power-cap floor leaves too little range to characterize the law; Section 9). **Measurement.** Each card was measured on its own AWS EC2 instance (g6 for the L4, g5 for the A10G, a p4d on which the eight A100s ran disjoint slices of the model set in parallel, one model per GPU, and a g4dn for the T4), running a PyTorch Deep Learning AMI. Power caps were set with nvidia-smi -pl; board power was sampled from NVML at 20 Hz over a two-second window per operating point, and throughput R is iterations completed over the same window, where one iteration is one forward pass of the whole batch (a batch step). We report R in these batch steps per second throughout; per-sample rates scale by the batch size, which is held fixed within any single response curve. Within a response curve the caps were stepped in ascending order, each held for a 1.5-second settle before its two-second window with no separate warm-up pass, so the first window after a step slightly under-reads power while the card settles (Section 9).

**Schema and scope.** Three sweep collections and one trace make up ELF; a sweep row is one measured operating point (model, GPU, actuator level, batch, repetition). The batch-32 collection sweeps all four GPUs (two repetitions) and records, beyond power and throughput, a four-phase latency split (model load, host-to-device, compute, device-to-host). The batch collection records board power, GPU utilization, and throughput at a light-load baseline of batch 32 (mode *control* in the release) and at an auto-calibrated saturating batch that maximizes uncapped draw (mode *saturate*); every operating point is measured three times, and the reported curves average the repetitions. A third, clock-sweep collection locks the graphics clock (nvidia-smi -lgc, two repetitions) to reach operating points below the power-cap floor on the L4 and, at a matched batch on the A10G, to compare the two actuators; it supplies Figures 1, 3, and 4. The trace (about 50 Hz) captures power and latency while the cap is stepped, for the actuation analysis of Section 7. In total ELF holds about 5500 measured rows across the sweep collections plus the actuation trace. The dataset, this schema in full, and the exact measurement scripts are released under CC-BY-4.0 and archived with a DOI (see Data availability); Section 9 states exactly what the numbers do and do not cover.

**Controlling the energy spent.** Energy per inference is set by the GPU's operating point, and the driver exposes two ways to move it. The *power cap* (nvidia-smi -pl) bounds board watts directly and lets the card pick the voltage and frequency that fit; the *graphics-clock lock* (nvidia-smi -lgc) fixes the frequency and lets power float. Both are DVFS underneath: a lower operating point means a lower voltage and less power. We sweep the power cap as the primary control and add a clock sweep on the cards whose cap floor is too high to reach the energy minimum (Section 9). Where their draw ranges overlap the two are equivalent within the measured range: Figure 1 shows, on the A10G at a matched batch, that the power-cap and clock sweeps trace the same energy-per-inference curve, and the clock simply reaches below the power cap's driver floor. The object we control is therefore the energy operating point, not the knob; we report board draw and energy throughout and treat which actuator reaches a point as an implementation detail.

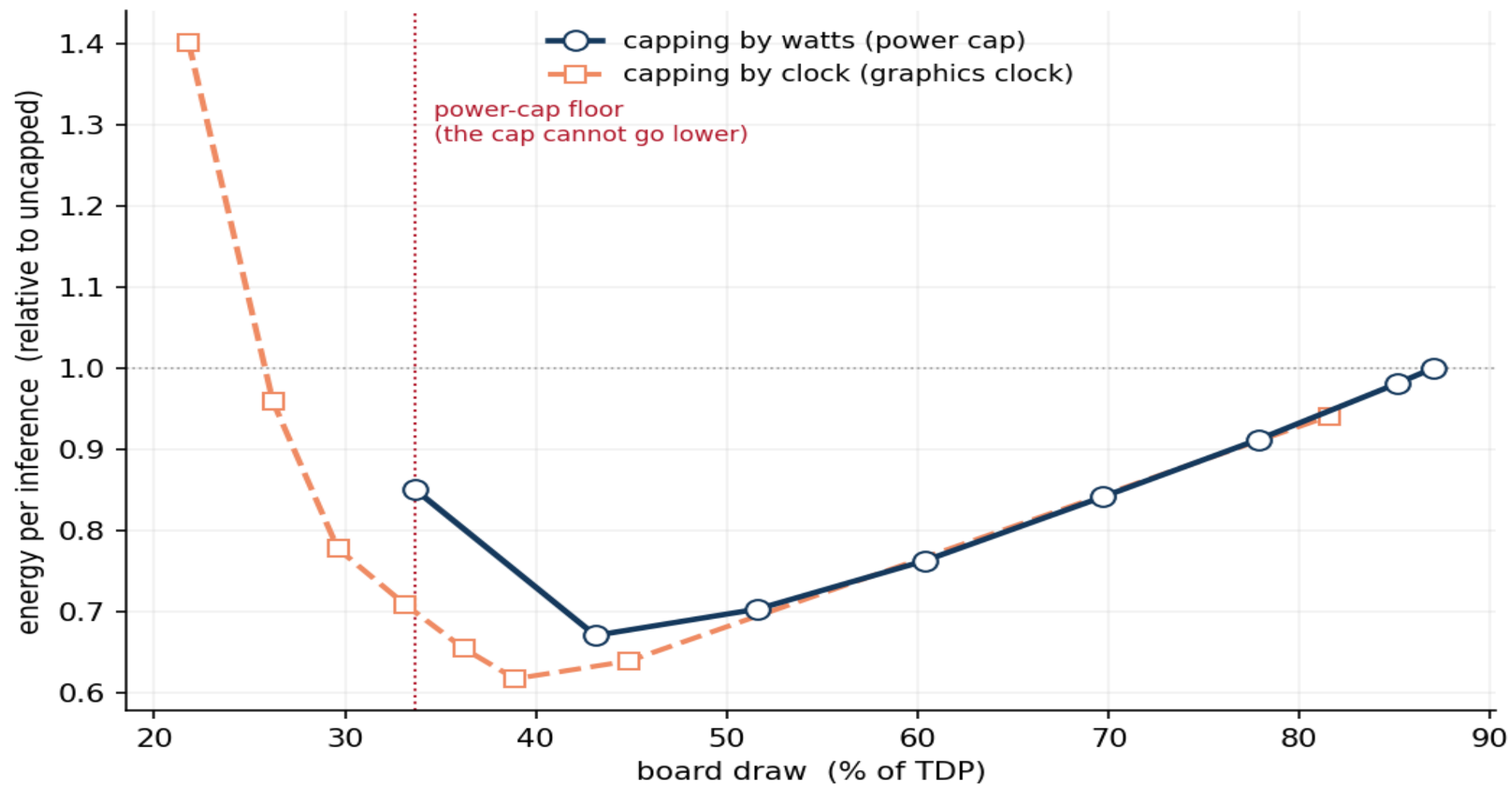

***Figure 1.*** *Watts or clock, the same energy curve. On the A10G (convnext_small, batch 512) the power-cap sweep (navy) and the graphics-clock sweep (orange) give the same energy per inference where their board-draw ranges overlap (mean difference about 3 per cent), and the clock reaches below the power cap's driver floor (dotted line) to a lower draw. Energy is normalized to the uncapped value.*

## 5. The response surface and its law

**Where the law comes from.** The functional form is not assumed; it follows from how a GPU turns power into rate. A processor's dynamic power obeys the CMOS switching relation $P_{\mathrm{dyn}} \propto C_{\mathrm{sw}} f V^2$ (switched capacitance $C_{\mathrm{sw}}$, clock frequency $f$, supply voltage $V$ squared), on top of a static floor $P_0$ from leakage and fixed board draw. For a compute-bound inference kernel the achieved rate rises with the clock, $\theta \propto f$, and a higher clock needs a higher voltage for the transistors to switch in time, so along the card's frequency-scaling curve $V$ climbs with $f$. With the switched capacitance fixed, $P_{\mathrm{dyn}} \propto f V^2$: where voltage is flat this is linear in the rate, and where voltage tracks the clock it reaches and can exceed the cubic $f^3$, so $P_{\mathrm{dyn}} \propto \theta^\beta$ with $\beta \geq 1$ growing as the voltage-frequency curve steepens. Adding the floor gives the law we fit, $P(\theta) = P_0 + a\theta^\beta$, and it predicts what we then observe: the exponent $\beta$ is set by the card's $V$-$f$ characteristic, so it travels with the card rather than with the model.

Fitting $P(\theta) = P_0 + a\theta^\beta$ to each cap sweep (173 configurations of GPU, workload, and sweep collection) gives a median $R^2 = 0.986$, against 0.844 for a linear model (Figure 2). The law is not an artifact of light load: the fit is equally clean under a saturating batch (median $R^2$ 0.98) and at light load (median 0.99). The fitted exponents bear out the prediction that $\beta$ tracks the card more than the model (batch-32 medians L4 1.4, A100 5.2, A10G 8.0); we report it as a tendency rather than a constant: the within-card spread is wide, and 41 of 173 fits reach the $\beta = 8$ ceiling imposed on the fit, so the largest per-card medians are lower bounds and the upper tail of the exponent is not identifiable from these constrained fits.

The decisive property is that $P(\theta)$ is a law of the operating point, not of the rate. The counter-example is direct: on one GPU, two workloads delivering the same throughput differ by 2.7× in board power. The same holds within a single workload: the same throughput is reachable at different power through different (batch, cap) points, so pooling batch sizes and fitting power against rate alone collapses the fit to a median $R^2$ near zero. Were power a function of rate alone, hardware space would suffice to reason about energy; because it is a function of the operating point, energy space is required.

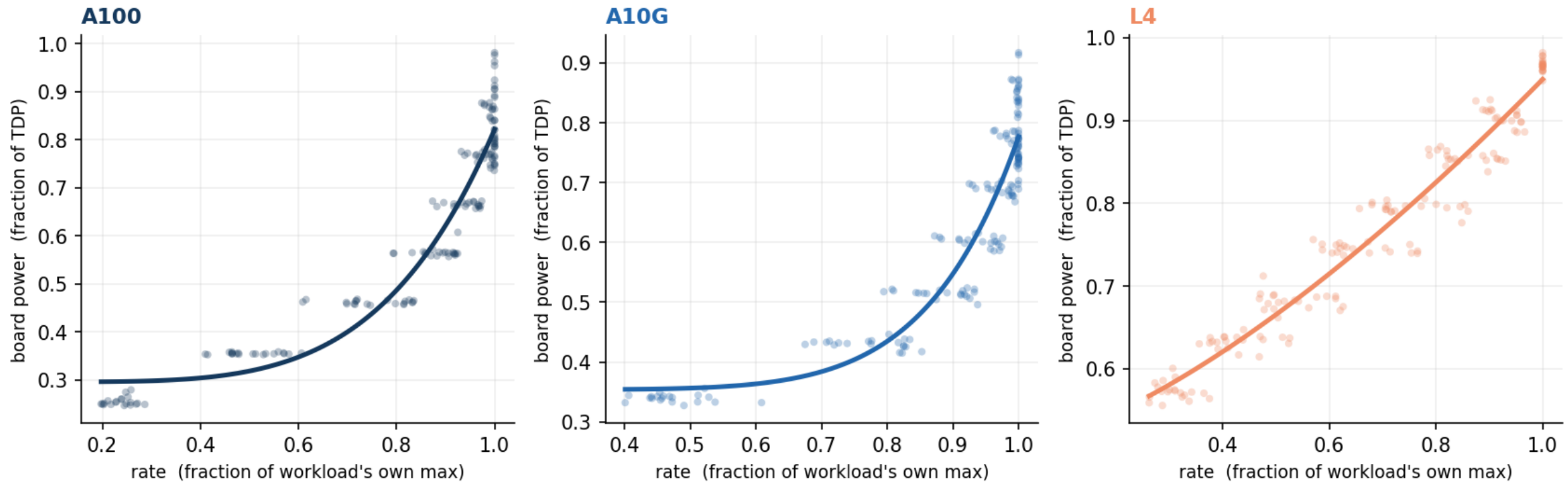

***Figure 2.*** *The response law P(θ)=P0+aθβ is a card property. One panel per GPU (A100, A10G, L4); each dot is a measured operating point for one of the 20 workloads in the loaded regime, with the rate normalized to that workload's own maximum and power to the card's TDP. All workloads closely follow a single fitted law per card (line), showing the exponent travels with the card, not the model: pooled fit R2 of 0.92 (A100), 0.88 (A10G), and 0.95 (L4), with β from about 1.5 on the L4 to 7.4 on the A10G.*

## 6. The Joule Point, and the price of reaching it

Because power is superlinear in rate while the floor $P_0$ is paid regardless, energy per inference $E = P/\theta$ is U-shaped (Figure 3): under-powering pays the floor over a slow job, over-powering pays the superlinear tail. We name the stationary point $dE/d\theta = 0$ the **Joule Point**, the per-job cap that minimizes energy per inference. On the large GPUs it sits well below TDP, at a median 46 per cent of TDP on the A100 and 43 per cent on the A10G, and capping to it cuts energy per inference by 29 per cent (A100) and 31 per cent (A10G). The well is deep on the A100 and A10G and shallow on the L4, where the minimum lies below the power cap's reachable floor and is exposed only by the graphics clock (Section 9).

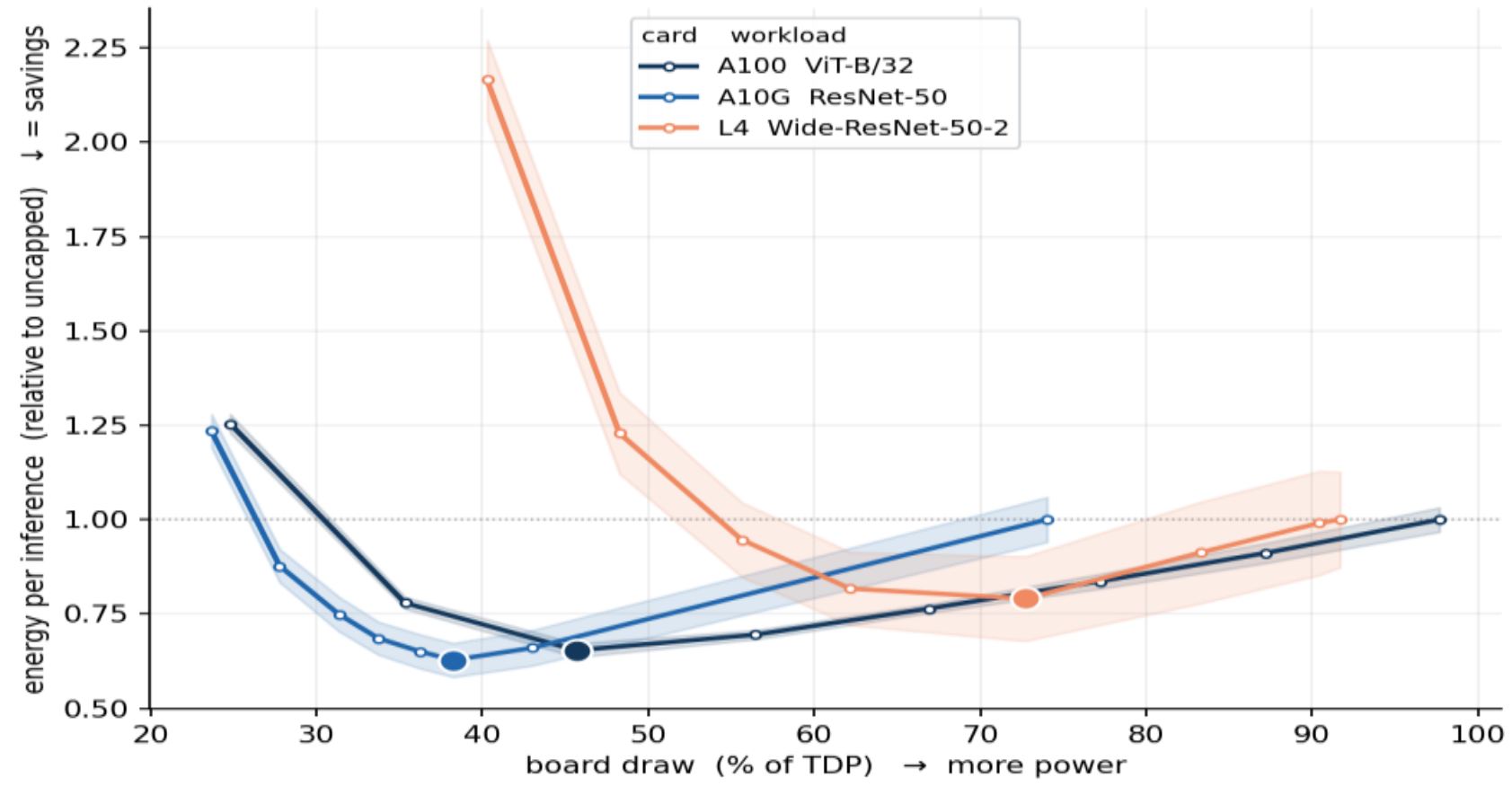

***Figure 3.*** *Every card has an energy U-shape, shown with the best-illustrating workload per card: energy per inference normalized to each card's own uncapped value (so every curve starts at 1.0 and dips by the energy saved), versus board draw as a fraction of TDP. The shaded band is one standard deviation over repetitions; the dot marks the Joule Point. Curves: ViT-B/32 on the A100, ResNet-50 on the A10G, and wide-ResNet-50 on the L4. The L4 reaches its interior minimum below the power-cap floor, so its well is exposed by lowering the graphics clock instead.*

**The Joule Point in closed form.** With the rate written normalized, $\theta \in (0,1]$, energy per inference is proportional to $P/\theta$ up to fixed per-job constants (the work per inference and the rate ceiling $R_{\max}$), which scale the energy but not the location of its minimum. Dividing the law by the rate splits energy into a floor term that falls with rate and a dynamic term that rises with it,

$e(\theta) = P(\theta)/\theta = P_0/\theta + a\theta^{\beta-1}$ (1)

Setting $de/d\theta = 0$ gives the energy-optimal rate, and substituting it back into the law gives the Joule-point power,

$\theta^* = \left(P_0/(a(\beta-1))\right)^{1/\beta}$ (2)

$P^* = P_0\,\beta/(\beta-1)$ (3)

The coefficient $a$ cancels: the Joule-point power depends only on the floor and the exponent. A minimum exists only for $\beta > 1$; real cards keep $\beta$ above one, so an energy minimum exists on every card we measure.

**The price of reaching it is exact.** Energy per inference is work-normalized, so serving the same aggregate throughput at the efficient cap by adding cards costs exactly the energy-per-step ratio; the card count cancels and the saving is real, not an accounting artifact. What it costs is capital and per-request latency, and these are the same number, because the card multiplier equals the latency ratio (both equal $1/\theta^*$, uncapped over capped throughput). The A100 saves 29 per cent energy for 1.24× cards and latency; the A10G, 31 per cent for 1.22× (Figure 4).

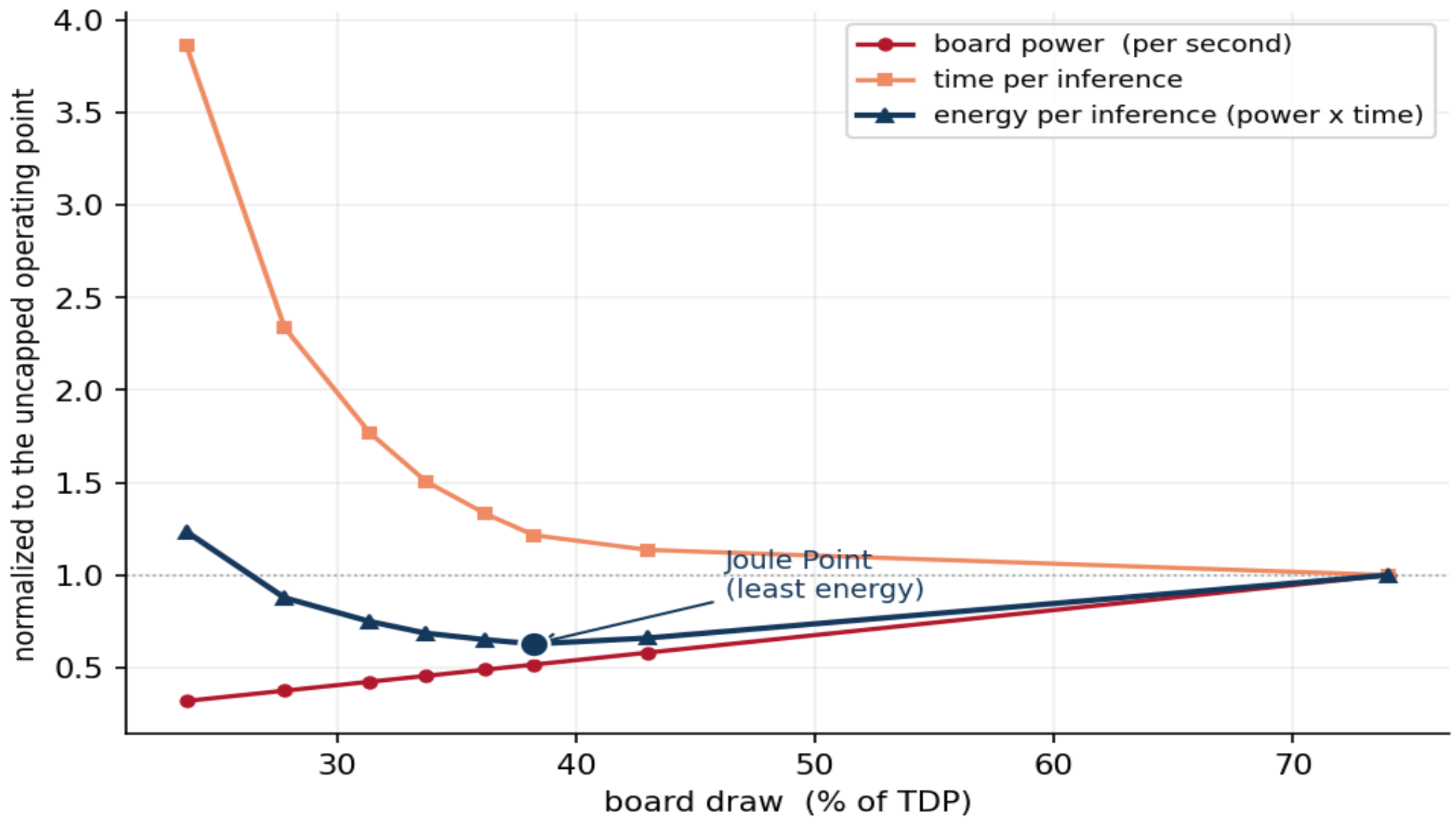

***Figure 4.*** *Why energy per inference is U-shaped, and what lowering the operating point costs (A10G, ResNet-50, via the graphics clock), each series normalized to its uncapped value. As the operating point is lowered, board power per second drops (red) but each inference takes longer (orange); their product, energy per inference (navy), is minimized in between at the Joule Point (dot). Left of the Joule Point a small further energy saving costs a large latency rise; right of it a steep energy cost buys little speed.*

**One fixed cap per card is nearly as efficient as per-model tuning.** Under load the Joule Point is nearly model-invariant, so a scheduler needs no per-job sweep or probe: a single per-card cap captures almost all the saving. On the A100 the energy-optimal cap sits near 46 per cent of TDP across the twenty workloads (standard deviation 4.7 points), and holding every model there costs a mean of just 0.9 per cent energy versus each model's own optimum (17 of 20 within two per cent). The A10G is tighter still: 43 per cent (standard deviation 2.8 points, penalty 0.4 per cent) (Table 1).

The cause is physical, not coincidence. Under a saturating batch the floor $P_0$ is the same for every workload, a hardware baseline: on the A100 the fitted floor is 120 W with a coefficient of variation of only 0.05 across the twenty models. Since the Joule-point power is $P^* = P_0\, \beta/(\beta - 1)$, set mainly by that floor, the optimum is a card property: the closed-form optima span only about 1.8 points of TDP on the A100 and 1.9 on the A10G. In a lighter, unsaturated regime the floor varies far more across models and one fixed cap would no longer suffice, so the invariance is a consequence of running loaded, the regime representative of throughput-oriented, well-batched serving.

***Table 1.*** *The Joule Point on the two large GPUs, under load: its location (per cent of TDP), the energy it saves per inference versus uncapped, the extra cards (equal to the latency ratio) needed to hold throughput, and the spread and penalty of using one fixed per-card cap across the twenty workloads.*

| card | Joule Point (%TDP) | energy saved (%) | cards = latency (×) | fixed-cap sd (pts) | fixed-cap penalty (%) |
|---|---|---|---|---|---|
| A100 | 46 | 29 | 1.24 | 4.7 | 0.9 |
| A10G | 43 | 31 | 1.22 | 2.8 | 0.4 |

Under load, one static cap per card, set once, has a mean energy penalty below one per cent versus each job's own optimum, so the online per-job search that prior energy-aware systems run is unnecessary in this regime.

**Board-level to node-level.** GPU board power is not node or facility power: a real node also draws host, memory, NIC, and power-supply overhead, and sits behind a facility PUE. Facility multipliers (PUE, PSU efficiency) are constant factors that cancel in the saving ratio, but a fixed per-GPU overhead does not, and it moves the optimum: folding a representative 100 to 200 W of per-GPU node overhead into the floor raises the node-level Joule Point to about 57 per cent of TDP on the A100 and 52 per cent on the A10G, and lowers the A100 saving to 12 to 18 per cent (the A10G, to 10 to 16 per cent). Capping remains a clear win once node overhead is counted, at a Joule Point a little above the board-level optimum; the deepest savings belong to nodes whose non-GPU overhead is smallest.

## 7. Scheduling in energy space: a fleet simulation

The per-job Joule Point and the fixed per-card cap are static characterizations; a real fleet serves a drifting mix of jobs under a shared power budget, so we test whether energy-space scheduling holds up in motion. To compare controllers on identical inputs, we run a trace-driven simulation: a closed-loop scheduler drives a virtual fleet of 15 GPUs (five each of L4, A10G, and A100) over a 200-tick trace whose job mix drifts between light and heavy models (each tick draws one job per GPU from a seeded generator whose heavy-model fraction oscillates over a 60-tick period), under fleet power budgets of 50 to 70 per cent of total TDP. The simulation is not a fresh benchmark; it replays the measured curves of Section 4, which supply ground-truth power and latency at every (GPU, model, cap), so the only thing varied is the scheduling decision. Each job's SLO is 1.25× its best achievable latency unless stated otherwise, and the trace is seeded, so every number reproduces exactly. Both a hardware-space controller (competent placement, run uncapped) and an energy-space controller (cap each admitted job to the least power meeting its SLO) admit jobs greedily by tightest deadline first and serve only those they can complete within SLO under the budget; the two share the same admission order (tightest SLO first) and placement rule (each job to the lowest-power GPU that meets its SLO and has free capacity), differing only in whether they cap. Capping to the least SLO-feasible power both cuts a job's energy against the uncapped baseline and frees fleet power to admit more work; it coincides with the Joule Point when the SLO is loose enough that the Joule Point is feasible, and sits below it when the SLO is looser still. The objective here is fleet goodput under a power budget, not per-job energy: below the Joule Point a job spends more energy per inference but releases power that admits another job, so the scheduler trades single-job efficiency for throughput. Where energy per inference is the sole objective, the energy-minimizing cap is instead $\max(P^*, P_{\mathrm{SLO}})$, never below the Joule Point.

The energy-space controller matches or exceeds the hardware-space controller's SLO-met throughput while spending 18 to 45 per cent less energy per served job (the range spans SLO slack from +10 to +100 per cent, at a 60 per cent power budget). This capping gain is a property of the operating point, not the hardware mix: on a homogeneous A100 fleet, where no cross-GPU routing is possible, it is undiminished (28 per cent less energy per served job at a 50 per cent budget), because it comes from each job's own operating point, not from routing across a diverse fleet. A separate, additive gain, admission headroom (serving more jobs for the same budget by freeing power), does depend on the spread of draw across jobs.

Figure 5 isolates that allocation question in its cleanest form, separate from the closed-loop trace: a static fleet of 18 identical A100s, one measured model per GPU, with the fleet power budget swept from 100 down to 40 per cent of the 7.2 kW total TDP, and the best per-GPU cap assignment computed exactly at each budget. On aggregate throughput the per-GPU assignment adds only a modest gain; when the objective is goodput the same freedom is decisive: at a 60 per cent budget a single uniform cap serves 6 of the 18 jobs within SLO, while the per-GPU assignment serves 16.

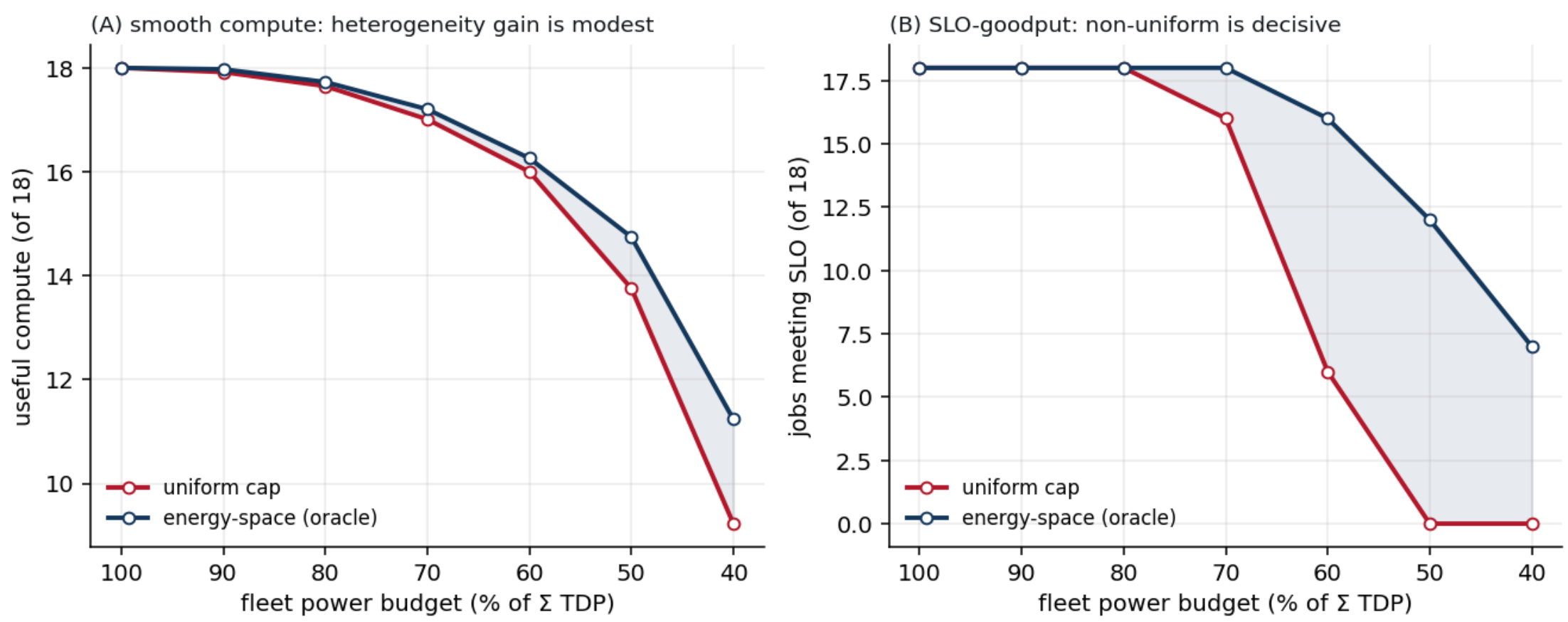

***Figure 5.*** *Compute-power frontier on a fleet of 18 identical A100s, one measured model per GPU, computed deterministically from the measured cap-sweep curves (repetitions averaged); this is the static allocation optimum, not the closed-loop trace. The x-axis sweeps the fleet power budget from 100 down to 40 per cent of the 7.2 kW total TDP. At each budget, energy-space allocation (each GPU its own cap, the exact optimum over the measured cap grid) is compared with the uniform baseline (one fleet-wide cap, the largest measured cap that fits the budget on all 18 GPUs). Left: useful compute, each model's throughput normalized to its own full-power peak and summed. Right: SLO goodput, the number of jobs whose latency stays within 10 per cent of that model's full-power latency. The advantage is modest for smooth throughput and decisive for goodput: as the budget tightens the uniform baseline collapses while the per-GPU assignment degrades gracefully. Both panels are computed on the saturating-batch sweeps, the loaded regime of Sections 5 to 7.*

The lever is also fast: stepping the power cap on a live A10G, board power settles to within five per cent of the new level in a median 191 ms, and rests at the job's natural draw when the cap is inert, so cap actuation is fast enough for second-scale control. As a concrete rule, ELF answers routing queries directly: for LLM decode, whenever the latency budget is about 158 ms or looser, the lowest-board-energy option is a single operating point, the A100 capped near 46 per cent of TDP, which meets that budget and is the most efficient choice in the fleet; running the A100 uncapped or falling back to a smaller card is a visible error. A binding cap also holds draw steady: uncapped A10G draw drifts by about 17 per cent as the card warms, and a cap makes power predictable.

## 8. Implications for AI data center operators

The following read directly off our measurements on the workloads and GPUs in ELF; they are prompts for evaluation, not deployment prescriptions. Real decisions turn on factors we do not model here, arrival patterns, production serving stacks, newer hardware, SLA and capacity economics, and carbon and power contracts, so each point below is a hypothesis to test against a given fleet.

- *A single static per-card cap may replace the per-job control loop, and is worth weighing against idling*. The energy-optimal cap moved little across our twenty workloads under load (Section 6), so the online, per-job search that energy-aware controllers run was unnecessary on every workload we measured; online control keeps its value in phase-varying or bursty regimes where the floor moves. On the operating points we measured, holding each GPU at that cap served the same work at roughly a quarter to a third less board energy (Section 6), the fleet controller spent 18 to 45 per cent less energy per served job (Section 7), and the cap settled in about 191 ms; whether that beats dropping or queueing work under a fluctuating budget is an experiment worth running.
- *Whether capping costs cards or is effectively free is set by the fleet's regime (Figure 6).* Where a deployment is limited by the megawatts it can draw rather than the cards it can buy, the Joule Point is a capacity lever, not only an energy one: under a fixed power envelope, running each GPU at its energy-optimal point delivers more useful work per watt, so efficiency becomes throughput and the case for capping stops depending on the price of energy at all. Behind a 1 MW feed the measured A100 optimum sits near 5,500 cards and serves about 50 per cent more inferences per megawatt than running that power uncapped (Figure 6). The same holds under light or bursty demand: when the average request is small relative to the card (few FLOPs per inference) or arrivals leave the fleet below capacity, holding aggregate throughput needs no extra cards, so capping removes energy at no capital cost, though request latency still rises. The capex penalty of capping appears only once demand saturates the fleet; below saturation the energy saving stands alone, opposed only by the split incentive, and running near TDP is right only where power is plentiful and capital scarce.
- *Sizing by energy per unit of work, not by peak TDP, exposes headroom the TDP number hides*. The efficient operating power sat near 43 to 46 per cent of rated TDP at the board level, and higher once node overhead is counted (Section 6); the roughly 1.24× extra cards needed to hold throughput is the explicit price to weigh, and whether it is worth paying depends on the price of energy relative to capital. How hard to cap depends on what latency is worth: keeping 95 per cent of a throughput utility (value that rises with each unit of throughput) on the A100 saved a median 13 per cent of energy, but 28 per cent under a deadline utility (value earned only for meeting the deadline), where finishing early buys nothing and the job can be capped harder for free.

- *Pricing that ignores power pulls the wrong way, so realigning it is the main lever on the incentive to cap.* Cloud GPUs are billed by the hour, not the watt, so a renter minimizes cost by running uncapped, the energy-worst point, while the party that would gain from capping (the operator's power bill, the grid) does not pay the rent. A per-joule surcharge is too weak to close this: at $80 per MWh a $12,000 A100 must run about 43 years for its electricity to equal its purchase price, so the energy line-item is dwarfed by the card-hours. The levers that bite are structural: first-party operators pay both capex and energy and should target the Joule Point directly; carbon pricing or power-constrained allocation raises the value of a watt; and a discounted capped tier gives tenants a reason to run capped.

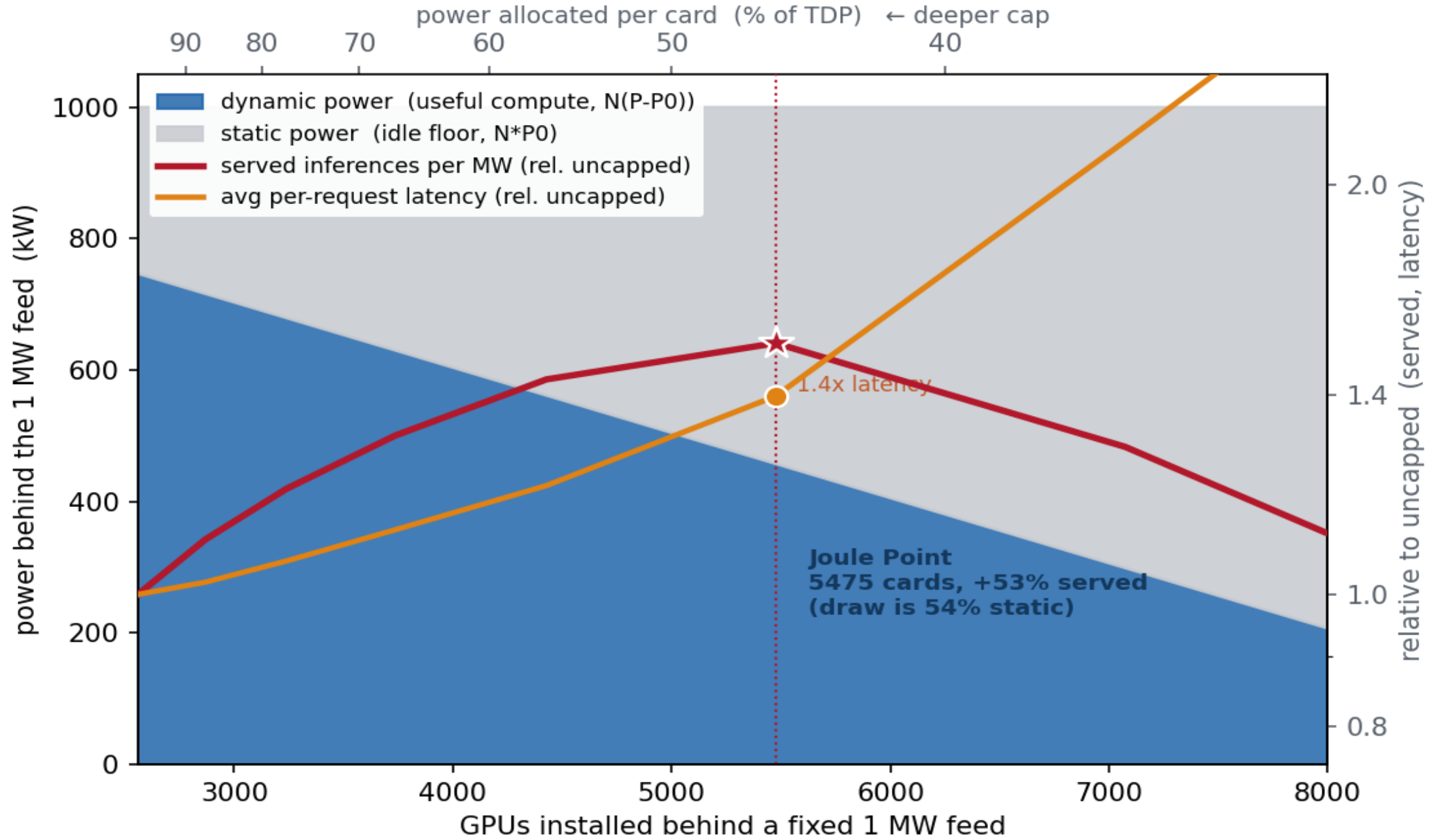

***Figure 6.*** *When power, not capital, is the binding constraint, the Joule Point is the interior optimum. Behind a fixed 1 MW feed (illustrated on the A100 running ViT-B/32 at batch 64), installing more GPUs forces each to a deeper cap (top axis), shifting the megawatt from dynamic, useful compute (blue) to the static idle floor (grey, taken as the least measured board draw). Served inferences per MW (red) peak at the Joule Point, near 5,500 cards, about 50 per cent above running the same megawatt uncapped; installing more cards is strictly dominated, serving fewer inferences at higher per-request latency (orange, right axis, log scale). Both served throughput and latency are relative to the uncapped operating point (= 1.0, the 100 per cent-TDP reference). This is the power-limited, demand-saturated regime; where demand is limited instead, one installs only enough cards to meet it and caps to the least SLO-feasible power (Section 7).*

## 9. Limitations

What this paper establishes, it establishes by direct measurement: a superlinear power-response law across 20 models on three NVIDIA inference GPUs, a Joule Point that under load is a per-card constant, and a fleet result driven end to end by measured curves. These findings hold for the workloads, cards, and loaded regime we swept, one model per GPU at fixed batch and precision, and that scope maps directly onto the extensions we set out below. The closed-form Joule Point explains the physics, why an energy minimum exists and why the shared floor $P_0$ makes it a per-card constant, while the deployed cap is set directly from the measured sweep; on the high-exponent cards the analytic optimum lands within about ten percentage points of TDP of that measured cap, and since the deployed cap uses the measurement the gap costs nothing. The cross-card relationships are scaling observations over three GPUs, so a fitted transfer model to unmeasured cards is future work, and the cap-sweep protocol that would build it carries over unchanged. That protocol extends directly to more workloads, more GPU generations and vendors, and a production serving engine such as vLLM or TensorRT-LLM (Section 10), where co-location, iteration-level batching, and a shifting prefill/decode mix would be exercised under live load. The economic analysis uses stated illustrative prices, card cost, electricity, and a four-year life, so its break-even figures scale transparently with those inputs.

The measurement scope is equally definite. Each load-regime operating point is a three-repetition sweep (the batch-32 collection uses two), so conclusions rest on whole-sweep trends rather than single cells, and every headline claim is of that kind. The first two-second window after a cap change under-reads board power while the card settles, which biases the reported savings downward; the numbers are a lower bound. Setting a cap requires nvidia-smi privileges, which bare-metal and VM tenants hold, the tenancy tier where large fleets run; the fleet evaluation of Section 7 is trace-driven over measured curves, so a live deployment would test the law under production batching and arrival variation. Finally, as Section 4 and Figure 1 establish, the subject is the energy-optimal operating point and the knob is a detail: on the large GPUs the power cap reaches that point directly; on the L4 the cap floor (about 56 per cent of TDP) sits above it, so the locked clock reaches it instead (median 8 per cent saving); and the T4's cap range is too narrow to characterize the law, so it is released in ELF and held out of the quantitative analysis.

## 10. Conclusion and future work

Hardware-space scheduling allocates GPUs; the cost that matters is joules, and joules are set by the operating point, not the device. Measuring that surface for 20 models on three GPUs gave a superlinear power-response law (median $R^2$ of 0.99), and the law gave a Joule Point: a cap at 43 to 46 per cent of TDP on the large cards that cuts energy per inference by a quarter to a third while each request runs about 1.2 times slower, a cost that is exact because the card multiplier to hold throughput equals the latency ratio. Under load the Joule Point is nearly a per-card constant, so a single static cap set once captures what prior systems search for per job at runtime, at a mean penalty under one per cent; a fleet controller replaying measured curves serves equal work for 18 to 45 per cent less energy, and behind a fixed power feed the same cap becomes a capacity lever, maximizing inferences served per megawatt. The physics is favorable and the actuation is a one-line command; what stands in the way is per-hour pricing, which rewards the renter for running at the energy-worst point. That obstacle is economic, not technical, and Section 8 lays out the structural remedies that would remove it.

Two extensions follow directly. The measured cross-card regularities rest on three GPUs, so the first is a fitted transfer model that places the Joule Point on unmeasured cards, built by running the unchanged cap-sweep protocol on more generations and vendors. The second is validation under a production serving engine such as vLLM or TensorRT-LLM, where iteration-level batching and a shifting prefill/decode mix would test whether the law persists under live load; bursty regimes, where the power floor moves, would also test whether the static cap needs an online supervisor. Orthogonally, adaptive-inference methods, early exit [5, 11], conditional computation [6, 7], and speculative decoding [8], vary the computation per request while the cap varies the power of that computation; the two knobs compose in the same coordinate system, and measuring their joint surface is a natural sequel.

### Data and code availability

The ELF dataset, twenty inference models swept over the board power cap on four NVIDIA GPUs, together with the exact AWS measurement harnesses that produced it, is archived at Zenodo, doi:10.5281/zenodo.22058568, under CC-BY-4.0 (data) and MIT (code). Every empirical figure and measurement-derived number in this paper is computed from that dataset by the accompanying build script.